\documentclass[letterpaper,10pt]{article}
\usepackage[left=1in,right=1in,top=1in,bottom=1in]{geometry}
\usepackage[utf8]{inputenc}
\usepackage{amsthm}
\usepackage{amsmath, amssymb}
\usepackage{enumitem}
\usepackage{algorithm}
\usepackage{soul}
\usepackage{algpseudocode}
\usepackage{float}
\usepackage{color}
\usepackage{tabularx}
\usepackage{hyperref}
\usepackage{url}
\usepackage{breakurl}
\usepackage{xifthen}
\usepackage{tikz}

\newcommand{\Cmnt}[2][]{%
     \ifthenelse{\isempty{#1}}{\Comment{#2}}%
         {\Comment{\parbox[t]{#1\linewidth}}{#2}}}

\algblockdefx[Struct]{Struct}{EndStruct}%
  [1]{struct #1 }%
  {end struct}

\newcommand{\remove}[1]{}

\floatname{algorithm}{Pseudocode} 
\newlength\mylen
\begin{document}

\title{RCanopus: A Geo-distributed, Scalable, Byzantine Fault-Tolerant Consensus Protocol}
\author{S. Keshav, W. Golab\footnote{W. Golab is with the Department of Electrical and Computer Engineering at the University of Waterloo}, B. Wong, S. Rizvi, and S. Gorbunov \\School of Computer Science, University of Waterloo\\ Waterloo, Ontario, Canada\\\{keshav, wgolab, bernard, sm3rizvi, sgorbunov\}@uwaterloo.ca}
\maketitle
\date
\begin{flushright}\normalsize{v1.20} \end{flushright}

\section{Introduction}
Byzantine-fault tolerant (BFT) distributed consensus is a key enabler for many distributed systems including
distributed databases and blockchains~\cite{bano2017consensus}. 
Existing BFT consensus algorithms have poor scalability due
the need for all-to-all communication between participants.
Thus, if there is a need for high transaction rate,
BFT consensus must be carried out only between a small number of 
participants, with the additional requirements that these participants be
geographically close, to mitigate latency from speed-of-light propagation delays.

The key idea in this work is to group participants into multiple geographically-close
Byzantine Fault Tolerant groups (BGs). Assuming that each BG is fault tolerant,
they can achieve consensus relatively quickly and in parallel.
Then, the results from these groups can be merged using an globally-distributed
crash fault-tolerant consensus protocol, such as the recently proposed
Canopus~\cite{rizvi2017canopus} scalable consensus protocol, which uses
pipelining, batching, and communication on a tree-like overlay network
to achieve very high throughput in a geo-distributed setting.

In this paper, we describe \textit{RCanopus} (`resilent Canopus')
which can be viewed as extending Canopus to add \textit{liveness}, that is,
to recover from failure despite using unreliable failure detectors,
and to tolerate Byzantine attacks. 
It guarantees \textit{safety} even
in the presence of Byzantine attacks and network partitioning.

Our design relies on several key ideas:
\begin{itemize}
\item Nodes are \textit{geographically grouped}. The first level of nodes are grouped by server rack into superleafs (SLs),
allowing fast intra-group communication. At the second level, nodes from SLs in
each geographical region form a
\textit{Byzantine Group (BG)}. Finally, BGs are hierachically clustered and execute the Canopus protocol in parallel.
\item BGs let RCanopus
achieve BFT consensus in parallel at relatively high throughputs
\item To deal with high-latency inter-datacenter links, consensus cycles are \textit{pipelined}, which allows multiple consensus cycles to be executed in parallel, increasing throughput

\end{itemize}

The remainder of the paper is structured as follows. 
In Section \ref{sec:bg}, we outline the Canopus protocol, deferring details to Reference~\cite{rizvi2017canopus}. 
Section \ref{sec:ass} states our assumptions. We lay out some building blocks for
RCanopus in Section \ref{sec:bb} and Section \ref{sec:cat} discusses how we
categorize different types of faults. Subsequent sections
address each category of fault and Section \ref{sec:summary} summarizes
the mitigation mechanisms. 

\section{Background} 
\label{sec:bg}

Canopus is a distributed coordination protocol implemented using a globally distributed set of servers called
\textit{nodes}, each of which periodically collects a set of transactions,
called a \textit{transaction block} from its \textit{clients}.
The nodes execute a consensus protocol to decide a global order on their transaction
blocks.

Canopus does not rely on a single leader.
Instead, it uses a \textit{virtual tree overlay} for message dissemination 
to limit network traffic across oversubscribed links. It leverages hardware redundancies, both
within a rack and inside the network fabric, to reduce both protocol complexity
and communication overhead. These design decisions enable Canopus to support
large deployments without significant performance degradation.

The Canopus protocol divides execution into a sequence of \textit{consensus cycles}. 
At the end of every consensus cycle, 
all nodes achieve \textit{agreement} or \textit{finality}, that is,
a total order on their inputs.\footnote{Some optimizations discussed later in 
this document postpone the commitment of transactions until a later cycle.}
Each cycle is labeled with a monotonically increasing cycle ID.
During a consensus cycle, the protocol determines the order of pending
write requests within transaction blocks received by nodes 
from clients before the start of the cycle and
performs the write requests in the same order at every node in the group. Read
requests are responded to by the node receiving it.
Canopus provides
linearizable consistency while allowing any node to service read requests and
without needing to disseminate read requests.

Canopus determines the ordering of the write requests by having each node, for
each cycle, independently choose a large random number, then ordering 
transaction blocks based on these random numbers. Ties are expected to be rare and are
broken deterministically using the unique IDs of the nodes.  Requests received
by the same node are ordered by their order of arrival, which maintains
request order for a client that sends multiple outstanding requests
in one session with a server during the same consensus cycle.
    
During each consensus cycle, each Canopus node disseminates the write requests
it receives during the previous cycle to every other node in a series of
\textit{rounds}.
Instead of directly broadcasting requests to every node in the group, which can
create significant strain on oversubscribed links in a datacenter network or
wide-area links in a multi-datacenter deployment, message dissemination follows
paths on a topology-aware virtual tree overlay.
Specifically, Canopus uses a Leaf-Only Tree overlay~\cite{allavena2006lot}, 
that allows nodes arranged in a logical tree to compute an
arbitrary global aggregation function. 
Each round computes the state (ordered set of transactions) at subsequently 
higher tiers of the overlay tree.

\section{Assumptions}
\label{sec:ass}

We now state the assumptions made in the design of RCanopus.

\begin{itemize}
\item \textbf{A1. Crash-stop failures:} Other than Byzantine failures (discussed as assumption A6), nodes fail by crashing: 
there are no transient failures. A failed node rejoins the system only through a node-join protocol (see Section \ref{sec:join}). In contrast, network failures can be transient, so that a network partition may recover due to extraneous recovery actions. 
%
%
%
%

\item \textbf{A2: Reliable communication channel:} We assume
the existence of non-Byzantine
communication channels between all pairs of nodes, and between clients and nodes,  that 
are not intermediated by other nodes, but only by tamper-proof routers and links. Hence, we assume that there are no message failures, such as losses, corruption, duplication, or reordering. In practice, this is easily achieved using a combination of TCP and dynamic network routing. 
%
%
\item \textbf{A3. Synchrony within an SL:} 
We assume that nodes within a SL, which, by definition, are connected by the same switch and in the same rack, run in a synchronous environment 
in which the communication and processing delays are bounded by a known value $\delta$. 
%
%
Formally, 
\begin{enumerate}
\item in the absence of failure, the maximum communication delays between the nodes are known and bounded
\item the maximum processing time required to execute each step of a deterministic algorithm is known and bounded
\end{enumerate}
and these bounds hold \textit{despite message, node, and link failures}.
Thus, we assume the existence of an atomic broadcast primitive \texttt{SLBroadcast(value)} that 
satisfies the following properties~\cite{modular}:
\begin{enumerate}
\item  \textbf{Validity}: If a correct process broadcasts a message $m$, then it eventually delivers $m$.
\item  \textbf{Agreement}: If a correct process delivers a message $m$, then all correct processes eventually deliver $m$.
\item \textbf{Integrity}: For any message $m$, every correct process delivers $m$ at most once, and only if $m$ was previously broadcast by its sender.
\item \textbf{Total order}: If correct processes $p$ and $q$ both deliver messages $m$ and $m'$, then $p$ delivers $m$ before $m'$ if and only if $q$ delivers $m$ before $m'$.
%
%
\end{enumerate}
Such a primitive can be built using an approach such as AllConcur~\cite{poke2017allconcur}.
Moreover, using an approach such as Zoolander~\cite{stewart2013zoolander},
we assume if a call is made by a node in some SL to \texttt{SLBroadcast(value)} at time $t$ then
it is received by all other live nodes in the same SL by time $t+\delta$.

\item \textbf{A4. Asynchronous inter-SL communication:} 
To make progress despite the FLP impossibility result\cite{fischer1985impossibility}, which states that 
safety and liveness cannot be simultaneously guaranteed
in asynchronous environments, 
our design gives up liveness in executions where the environment is unstable (e.g., messages are not delivered in a timely manner or processing of requests at servers is unusually slow).
A system that does not guarantee liveness in all executions is exempt from the FLP result, and is able to maintain consistency
despite (and during) network partitions.

Some parts of RCanopus, such as the global membership service, 
do require weak synchrony assumptions.  However, these components operate on long time scales and can therefore be configured with conservatively long timeouts for failure detection without compromising performance.

Practically speaking, this means that
in the absence of a network partition,
we assume that there is an upper bound $\Delta$ 
on the sum of inter-SL message delivery 
and server response times
such that it is possible for a node $n$
to send a message to another node $m$ and set a timeout such that, 
if the timeout expires, then $n$ can assume with high confidence
that $m$ is dead.
Though there
is some possibility that $m$ is actually only slow, not dead, or
alive, but on the other side of a network partition; in this case, the node
coordinates with other nodes to establish consensus on $m$'s status.
The timeout value should be chosen long enough to allow for transient
network partitions to recover; a partition lasting longer
than $\Delta$ is equivalent to a permanent partition\footnote{During a network
partition, nodes on one side of the partition are unable to
communicate with nodes on the other side. This does not make
the network asynchronous: in an asynchronous network,
any network communication may be subject to failure.}.
%
Moreover, we do not require clocks at different nodes to be
synchronized or run at the same rate, since the protocols
are self-synchronizing.

\item \textbf{A5: PKI:} We assume that every client and every Canopus node
has its own ID and own private/public key pair, so that every signed communication
is non-repudiable and non-falsifiable. The public key of a node can be viewed
as its unique node identifier. 

\item \textbf{A6: Byzantine failure of a super-leaf:} We conservatively assume that if even one node
in an SL suffers from a Byzantine failure, the entire SL is Byzantine.
This is based on the pragmatic observation that all the nodes in an SL
are likely to be homogeneous and in the same rack. 
Hence, if one of them has been maliciously taken over, it is very likely
the rest of the nodes on the rack have also been similarly compromised.

\item \textbf{A7: Byzantine groups:} We assume that we can partition
$\mathcal{S}$, the set of SLs,
into a number of mutually non-overlapping Byzantine groups (\textit{BGs}) $B_i$ where $\bigcup_i B_i = \mathcal{S}$
such that  
(a) sibling SLs in each Byzantine group $B_i$
are geographically `close' and 
(b) if there can be $f_i$ Byzantine failures in BG $i$,
then $|B_i| > 3f_i\ \ \forall i $.
In other words, each BG is resilient to Byzantine node failures.
Thus, the only failure mode for a BG is 
a network partition that prevents it from achieving quorum.
From outside the BG, this appears as an atomic failure of the entire BG.

Note that to prevent Byzantine faults, all communications that are sent on behalf of the BG $B_i$
must be signed with a \textit{quorum certificate} which proves that $2f_i+1$
members of the BG (including at least $f_i+1$ correct nodes) agree on the communication. 

In practice, we expect each member of a Byzantine group to be hosted by a different cloud service provider. Thus, a Byzantine failure models a security breach in a cloud service provider. We expect that in most practical cases there will be only one such breach ongoing, so that,  $\forall i, f_i=1$.

RCanopus does not provide safety if there are more than $f$
malicious nodes in a single BG.


\item \textbf{A8: BG leaders form a Byzantine fault-tolerant group:} Each BG elects a \textit{BG leader} that
participates in a BFT consensus protocol. We assume that if the number of BG leaders who
are subject to Byzantine failure is at most $f_g$ then 
the number of BGs exceeds $3f_g$.

\end{itemize}

\section{Building blocks}
\label{sec:bb}
This section describes the major building blocks in our design.

\subsection{Intra-SuperLeaf algorithms} 
\label{sec:islbb}

Based on the assumption of a synchronous environment within a SL,
we use standard approaches for crash-fault tolerant (CFT)
\textit{group membership},  
\textit{leader election}, 
and \textit{atomic broadcast} 
protocols.
\begin{itemize}
    \item A \textit{group membership} algorithm allows a set of nodes participating in 
a distributed protocol
to learn of each other. 
When a node leaves, the remaining nodes agree that the node has
left, and when a node joins, all other nodes agree that the new node is
a member.
    \item \textit{Leader election} involves agreeing on the identity of a distinguished node 
from a set of eligible nodes.  In practice, multiple rounds of leader election
may need to executed before a leader is chosen
because a tentatively-chosen leader may fail during the election process and may need to be replaced.
    \item With \textit{atomic broadcast}, meaning that all nodes can broadcast messages
to each other and receive messages in the same (arbitrarily decided) order. 
\end{itemize}
All three algorithms can be implemented using well-known systems such as
ZooKeeper~\cite{hunt2010zookeeper}, Raft~\cite{ongaro2014search} or AllConcur~\cite{poke2017allconcur}.
Hence, we do not discuss these further. 

\subsection{Byzantine groups}
\label{sec:bgs}
By Assumption A7, geographically-close SLs are grouped together to form a Byzantine Group (BG).
We expect these SLs to be hosted at a variety of hosting providers, so that
compromise of a single SL in a BG would not result in the compromise of its peer SLs
in the same BG.

We make each $BG_i$ resilient to Byzantine faults in its member SLs by relying on
existing Byzantine consensus protocols
such as PBFT~\cite{castro2002practical},
BFT-SMART~\cite{bessani2014state}, or
SBFT~\cite{golan2018sbft}.
Such a scheme is immune to up to $f_i$ simultaneous node failures as long as the number
of nodes in the BG is at least $3i+1$.
Importantly, each of these schemes delivers a
\textit{quorum certificate} that guarantees that the consensus 
value is valid.
The certificate contains signatures from a $(2f_i+1)$ quorum of nodes
that can be verified by the recipient and guarantees that the information 
contained in any message is valid.
Also, to prevent replay attacks, the certificate contains the current cycle number.

BFT consensus is also used to maintain SL membership status in the BG:
it is the set of SLs reporting transactions in the
latest completed BG consensus. If there is a network partition, 
SLs in the minority partition cannot submit their transactions, so are automatically 
excluded from membership.

\subsection{Global coordination}
\label{sec:gms}
In addition to consensus on membership within an SL and in a BG, 
we also need to achieve BFT consensus on a system-wide (`global') scale for:
\begin{itemize}
    \item obtaining of the set of emulators of a vnode at the BG level or higher
    \item dealing with apparent failures of BGs/vnodes arising from a network partition
    \item learning the quorum size in each BG (the number of signatures in each Byzantine Group's
quorum certificate that is necessary to validate it) during each consensus cycle
\item creating and storing \textit{global quorum certificates} to certify the set of BGs participating in each cycle (these are discussed in more detail in Section \ref{sec:resynch})
\end{itemize}

To do so, we deploy a \textit{global BFT membership service}
provided by a set of \textit{BG leaders},
each elected from among the monitors of every live BG.
Note that this distributed service itself needs to be BFT since BG leaders may suffer from Byzantine failures. 
Thus, this service is likely to have high latency 
in performing updates. Nevertheless, it can serve read requests 
relatively quickly since each BG leader participating
in this service can cache membership state along with its quorum certificate.

\subsubsection{Convergence Module-based global coordination}
\label{sec:cm}
An alternative approach is for 
global coordination to be done ``in-band'' with respect to the consensus protocol using a
specialized transaction type that must be endorsed by an authorized system administrator.
This approach associates group membership data with each cycle in a precise way: 
a group membership change proposed in cycle $C$ becomes effective in cycle $C+k$ where $k$ is the pre-defined and constant depth of the processing pipeline.

Convergence of each consensus cycle despite network partitions and BG failures is achieved using a service called the Convergence Module (CM) 
(details can be found in Appendix \ref{sec:cmapp}.
The CM internally keeps track of BGs that are suspected of being faulty, but operates orthogonally to the mechanism responsible for recording group membership.
In particular, a BG can participate in the consensus protocol and yet its input may be excluded from a particular consensus cycle by the CM because the BG was deemed faulty.
Repeated exclusions of a BG from consecutive consensus cycles may nevertheless prompt administrators to apply manual group membership changes.

Roughly speaking, the protocol functions as follows: a BG that stalls 
during a particular consensus cycle because it is unable to retrieve the 
inputs of one or more other BGs reports the situation to the CM, 
who then determines the output of the cycle under consideration as 
the union of the inputs of a carefully selected subset of BGs.
For example, the subset can be determined in a manner that ensures sufficient replication,
meaning that a BG's input is included in the output of a given cycle only if it has
been replicated at $R$ or more BGs for some administrator-defined threshold $R$.
To reduce overhead, optimizations are defined to bypass the CM entirely in absence of failures, and also to minimize interaction with the CM during network partitions that last many consensus cycles.
These optimizations deal with complex scenarios involving Byzantine failures and concurrency.

\section{Design}
\begin{figure}[!h]
    \includegraphics[width=\linewidth]{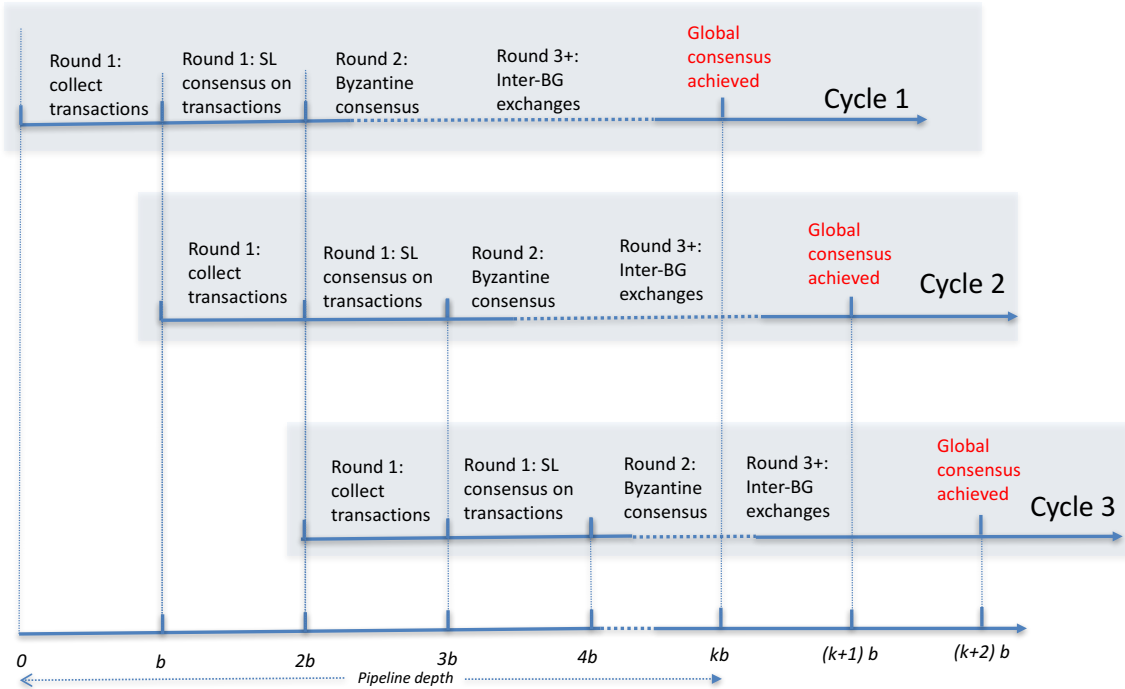}
	\caption{RCanopus pipeline. $b$ is chosen to be longer than the intra-SL consensus time. The long latency for inter-BG exchanges results in multiple simultaneous incomplete cycles  outstanding at any point in time.}
	\label{fig:pipeline}
\end{figure}

We now describe the design of the RCanopus system. We begin with actions taken by a \textbf{client}, which
sends transactions to RCanopus \textbf{nodes}. We then discuss operation of a node during each RCanopus
cycle. Some nodes play the role of a \textbf{representative} or \textbf{monitor}, and some monitors are also \textbf{BG leaders}. 
We discuss the actions taken by each such role.
Note that although we discuss actions in each cycle independently,
RCanopus is pipelined, that is, multiple cycles are in progress simultaneously, as shown in Figure \ref{fig:pipeline}.

\subsection{Client actions}
Clients\footnote{In the context of Hyperledger Fabric, the client is 
the orderer's front end, that accesses the RCanopus ordering service through the 
\textit{chain interface}.} send their transactions, each accompanied by 
a client-generated transaction id and a client-generated nonce, both signed by the 
the client's private key, to RCanopus nodes.
As in PBFT~\cite{castro2002practical}, clients first send 
transactions to a node in the closest SL. 
To prevent Byzantine nodes from simply ignoring their transactions,
if the client's transaction does not appear in the ordering service's output
within a configurable timeout period,
the request
is sent to nodes in $f_i$ additional, different SLs in $BG_i$.
Note that this requires clients to know the identities of the BG leaders and their public keys so that
the quorum size of $BG_i$ is known and verifiable.

A Byzantine client could try to confuse the system by sending inconsistent transactions
to nodes in different SLs.
This is,  however, easy to detect because a valid client must send the same transaction
to all nodes with the same client-generated nonce. 
Any deviation from this immediately identifies the client
as being malicious.

When they are ready to commit results from a cycle, 
nodes respond to clients with proof that their transactions
were incorporated into the global transaction order. The proof consists of two items\footnote{The proof does \textit{not} need to 
include a global quorum certificate (discussed later) showing the
certifying $BG_i$ was a member of the system during cycle $c$. This is because $BG_i$ can 
only generate and return a valid quorum certificate if
it is live (see Section \ref{sec:partition}) 
and has $f_i$ or fewer malicious nodes.}:
\begin{enumerate}
\item A Merkle tree branch connecting the client's transaction to the
      root for a transaction block (\textit{TB}).
\item A quorum certificate from $BG_i$ certifying that \textit{TB} was committed at cycle \textit{c}. 
Specifically, the certificate should include \textit{TB}'s Merkle tree root.	          
\end{enumerate}

\subsection{Node actions}
We now discuss the actions taken by all RCanopus nodes, independent of their role.

\subsubsection{Collecting client transactions}
In the first round of each consensus cycle, 
all nodes within an SL share their state (i.e., set of client transactions) 
with each other. 
Let \textit{batch time, b} be a constant chosen such that
all BGs can come to consensus withing time $b$, and such that $b >> \delta$,  
where $\delta$ is the maximum round-trip communication delay in the synchronous SL environment. 
Then, the first round lasts $2b$ seconds (see Figure \ref{fig:pipeline}).

During the first $b$ seconds of the round,
a node collects transactions from its clients.
It atomically broadcasts these transactions to its peers at the end of this time.
By Assumption A3, this state stabilizes within $b$ more seconds.
Hence, when the round ends,
all live nodes agree on the ordered list of transactions received
by nodes in the SL during the first half of the round\footnote{Note that some of these transactions may originate
 from servers that failed after communicating their state to peers in the SL using atomic broadcast. 
 This is silently ignored since it does not affect safety.}.
Because of pipelining, consensus cycles start every $b$ seconds as shown
in Figure \ref{fig:pipeline}.

\subsubsection{Transaction numbering}

A node assigns a transaction number to each transaction block (TB) of client transactions 
that it receives. This is used to identify
the position of the block in the global order.
It is possible for a Byzantine node to deliberately choose this number in a way 
that increases the chance for the TB ending up near the top or the bottom of the
global order. To prevent this attack, the transaction number of a TB
chosen as a Merkle hash of the transactions. 
Moreover, when TBs are further aggregated, 
the newly-created parent node orders its child subtrees by their Merkle roots,
forming a unique and non-manipulable total order. 

\subsubsection{Delayed commit}
Nodes commit the
state\footnote{In the context of Fabric, this corresponds to sending a list of transactions
to the orderer through the chain interface.} computed at the end of cycle $c$ at the end of cycle $c+1$, 
i.e, delayed by $b$ seconds.
This is because at this time
it can be sure that every other node in the system also will either commit or 
not commit a given other BG's state, guaranteeing safety.
This is discussed in more detail in Section \ref{sec:earlyexit}

\subsection{Representative and emulator actions}
\label{sec:rep}

\subsubsection{Representative election}
A certain number of nodes in an SL
are elected as SL representatives to fetch remote state. 
For example, these could be the monitor node and any other node designated to
act as a representative by the monitor. 
In the third and subsequent RCanopus rounds of each cycle,
\textit{representatives} from each SL fetch vnode state from the 
vnode's \textit{emulators} (i.e. any subtended pnode).

\subsubsection{State request and response}
At the start of each cycle, monitors learn of the list of vnodes 
(at the BG level and higher),
the IP addresses of their emulators, and BG quorum sizes from 
the global membership service, specifically,
from the local BG leader, which is the closest service representative\footnote{If this BG leader is Byzantine, it may deny service, in which case the request can be sent to any other BG leader. This requires the list of BG leaders and their public keys
to be globally known and periodically refreshed.}. 
These are then disseminated to the representatives.
Subsequently, representatives send \textit{state requests} to these emulators to learn 
about the state of one of their ancestor vnodes. 

As with clients, to improve performance despite Byzantine emulator faults, a
representative first contacts a single emulator for each BG,
starting a \textit{retry timer} (on the order of $\Delta$) before sending
the state request message\footnote{This timer value can be estimated by measuring response times, using the same
approach as that used for TCP retransmission timers.}.
On a timeout, it contacts $f_i$ additional emulators from $f_i$ distinct additional SLs in $BG_i$.

On receiving a request, the emulator sends 
either a \textit{state response} with the requested
state, along with a quorum certificate, 
or a null response, indicating that it does not yet have the 
desired state; subsequently sending the full response when it does become available. 
On receiving a non-null response if 
the quorum certificate has more signatures than the underlying BG's 
quorum size, the representative broadcasts this response to all 
SL nodes. On receiving such a broadcast, other representatives trying to obtain a 
response from that BG abort.

There are two possible failure cases:
\begin{itemize}
\item  The emulator may send a null response.
If so, the emulator will send the response later, when it become available (transforming a pull to a push).
\item The emulator may fail or be inordinately delayed, causing expiry of the retry timer at the representative.
If so, the representative tries to contact each of the
vnode's other emulators in turn.
If all emulators for the vnode are inaccessible, the representative indicates, using an atomic
broacast, that it is stalled. 
\end{itemize}

In both cases, it is possible that while waiting for a response,
some other representative may have already 
obtained the vnode's state. If so, the representative aborts it own request and takes no further action.

\subsubsection{Recovery from representative failure}

The monitor keeps track of the status of representatives (i.e, whether they are live, stalled, or failed).
On the failure of a representative, the monitor chooses a new representative,
which takes on the responsibilities of the failed representative.
If all representatives are stalled in trying to obtain a response from a BG, indicating either a BG failure
or a system partition,
the monitor initiates recovery, as discussed in Section \ref{sec:syspart}.

\subsubsection{Emulator behaviour on loss of consensus within the SL}
The intra-SL consensus service becomes unavailable when it loses quorum.
If a node acting as an emulator detects that the service has lost quorum, 
it stalls, that is, stops responding to state requests. 
This is to prevent a `split brain' 
in case of an SL partition (see Section \ref{sec:slpartition}).
When service is restored, the nodes need to rejoin
using a rejoin process, discussed in Section \ref{sec:join}.

\subsection{Monitor actions}

The monitor is responsible for several distinct sets of actions, as discussed in
this section.

\subsubsection{Byzantine consensus on transaction order}

%
%
%
%

In the second round of each cycle,
monitors in the SLs of each $BG_i$ come to a consensus about transaction order.
Since entire SLs can be Byzantine, it is necessary to use a BFT consensus protocol
to achieve consensus.
At the end of this consensus, each monitor obtains a total ordering of client
transactions and a quorum certificate
that certifies this ordering with $2f_i+1$ signatures.
Monitors also learn about which other monitors have failed,
either due to crash failure of their SL, Byzantine failure of their SL, or BG partition.

Recall that clients may send their transactions to multiple
nodes in multiple SLs, which causes
duplicate transactions to enter the system.
Duplicate transactions are detected using the per-transaction client-generated nonce,
and must be removed during the computation of the BG quorum certificate.
It is possible for client-initiated transactions to be received
by different nodes at different times, and may hence be part of different
RCanopus cycles. To deal with this situation, during the BG consensus
protocol, monitors buffer a hash
of recently-submitted transactions, removing duplicate transactions as necessary.

\subsubsection{Identifying and removing failed nodes from the list of BG emulators}
Monitors are responsible for updating the global consensus service to 
remove failed nodes from their SL (detected using the intra-SL
consensus protocol)
from the list of emulators for the enclosing BG.
This requires a quorum certificate from the BG to prevent Byzantine monitors
from causing trouble. Hence, the monitor
first contacts its peer monitors in its BG and 
proposes to them that a specific node in its SL has failed.
Each monitor sets a timer and independently tries to contact the 
potentially failed node.
If a quorum of monitors agrees that the node has indeed failed, then they generate
a quorum certificate, which is then submitted to the BG leader
to update the global consensus service,
removing this node from the list of the BG's emulators. 

\subsubsection{Initiating a response to system partition}
\label{sec:syspart}
Monitors are responsible for initiating the response to a network partition.
Recall that at the start of each consensus cycle, the monitor from each 
SL obtains a list of emulators for each vnode at the BG level and above
from its BG leader, who is the local representative for the global consensus service.
If there is a system partition, BG leaders in the minority partition(s)
will (eventually) fail to achieve quorum
and the monitor will receive a `loss of quorum' message\footnote{Even 
if they do obtain a stale list of emulators, the SL's representatives 
will stall when trying to contact
these emulators, preserving safety.}. 
This results in the stalling of all nodes in all
BGs in the minority partition, as desired.
Correct nodes will rejoin the system only by using the node re-join process, 
when they are asked to rejoin by their BG leader (accompanied by a global quorum certificate).

We now consider the situation, when,
during a network partition,
representatives potentially obtain a stale list of emulators corresponding
to inaccessible BGs, stalling
representatives who are waiting for a response from them.
These stalled representatives indicate their status to their monitor. 
When the monitor
detects that all the representatives
in the SL have stalled waiting for state from a remote BG I, it initiates
a Byzantine consensus on the status of this BG with peer monitors.
On invocation, monitors in the BG try to achieve consensus on the state of the stalled BG.
Specifically, each monitor checks
whether or not its representatives think that the BG has become uncontactable.
If so (because all of its representatives were unable to contact the SL),
then it agrees to the proposal that the BG be marked as failed.

The outcome of the consensus is either agreement that the BG has failed,
or a quorum of nodes respond with the state of the BG thought to have failed,
which is then shared with the monitors in the BG.
If all the monitors in the BG agree that BG I is indeed inaccessible,
then the BG leader initiates consensus on this at the global level, as discussed 
in Section \ref{sec:bgl}.

\subsubsection{Recovery from monitor failure}
Nodes in each SL keep track of the status of the monitor.
On monitor failure, leader election is used to choose a new monitor,
which takes on the responsibilities of the failed monitor.
In particular, the new monitor replaces the old monitor as a 
member of the BG. It also chooses new representatives, if necessary.

\subsection{BG leader actions}
\label{sec:bgl}

In each BG, a leader election amongst the monitors
is used to elect a \textit{BG leader}.
The set of BG leaders is collectively responsible for providing the
global consensus service discussed in Section \ref{sec:gms}. 
The alternative CM service is discussed in Appendix B.

\subsubsection{Responding to membership requests}
BG leaders respond to BG membership requests from representatives
with a list of vnodes at the BG level and higher in the system and the
IP addresses of their emulators.
To prevent attacks, this list is released only after validation that access is permitted to the requestor. 
For each BG, the BG leader also responds with the size of its quorum,
so that it is possible to verify that a response from an emulator has the
requisite number of signatures. 
The response itself is signed by a quorum of BG leaders.

\subsubsection{Dealing with and recovering from a system partition}
\label{sec:partition}
When a BG comes to BFT consensus that some set of other BGs are unreachable,
its BG leader contacts its BG leader peers to 
initiate a BFT consensus on membership.
At the end of this consensus, BG leaders in the super-majority
know of the set of BGs unreachable from their partition,
and they inform the monitors in their own BGs (along with a quorum 
certificate) to not stall waiting for inputs from these inaccessible BGs.
This allows recovery of liveness. BG leaders in the minority partition(s),
however, stall, and this causes their BGs to stall, as desired.

This process is expensive, but is only resorted to rarely:
only in case of a BG failure or system partition.
Moreover, it works correctly even if a BG's failure detectors
are unreliable: what is important is not that a BG's failure
be perfectly known, but that the rest of the BGs agree to ignore it. 

To preserve safety, once emulators in some BG have been informed by their BG leader that 
some other BG has failed, 
they should not respond to requests from representatives in the
failed BG until either their BG leader informs them
that the formerly failed BG has re-joined using the periodic global BFT consensus,
described next.
Moreover, nodes in the BG do not respond to clients, also to preserve safety.

\subsubsection{Periodic resynchronization}
\label{sec:resynch}
BG leaders periodically establish consensus on the
global set of BGs and their emulators using a BFT consensus protocol.
Specifically, during each such global consensus, the BG leaders use a BFT consensus protocol to
agree on
(a) the list of BGs in the system (b) the set of emulators
for each BG, and (c) the quorum size for each BG.
Any updates to the list of BGs be signed by at least $2f_g+1$ BG leaders;
updates to the set of emulators and the quorum size of $BG_i$ must be signed by 
at least $2f_i +1$ monitors in that BG.

BGs and SLs are only allowed to become part of the system during this synchronization
point.
Thus, in the period between synchronizations, which we call an \textit{epoch}, 
BG membership is static, other than removal of BGs, subsequent to the detection
and consensus on a BG failure or system partition event. Moreover, the list of emulators for
each BG is also not allowed to change. 

One outcome of the synchronization is a \textit{global quorum certificate} 
per epoch that certifies the set of participating BGs for that epoch as well as the cycles that
have been committed during that epoch.
This must be signed by at least $2f_g+1$ BG leaders.
The global quorum certificate, along with a corresponding set of BG-specific quorum
certificates, fully describes the set of transactions in a cycle and the BGs
who participated in the cycle, and this description
is both non-manipulable and immutable.
Hence, 
this certificate can be replayed to newly joined BGs to resynchronize them. 

While synchronization process is expensive,
doing it once every, say, 100,000 cycles,
amortizes the cost\footnote{In practice, it is likely that BG additions will be rare. 
It may hence be sufficient to require BG additions to be done out-of-band, 
and manually, with periodic consensus only used for refreshing the list of 
BG emulators or dealing with system-wide network partitions.}  
Note that because the set of emulators for each BG is only updated
during the periodic global consensus,
the list of emulators could be stale, missing a newly live potential emulator
or leading to the global service responding
with an emulator that is actually not live.
These errors are benign, due to the use of multiple emulators for each vnode. 
If a live node is missed, other live nodes are available to respond to queries from representatives. 
Symmetrically, if a node thought to be live is actually dead, or behind a partition, this is identical to the case when the node dies after the membership response. So, this level of staleness does not pose a problem.

\subsection{Node/SL/BG Join/Re-join Protocol}
\label{sec:join}

A new or a newly-live node must first make its presence known 
to the intra-SL membership service.
If the node is elected as a leader, it becomes both a monitor
and a representative, and chooses some other nodes in the SL to 
serve as representatives. 

Special care must be taken when a node joins an SL that is thought by
its BG peer SLs to be dead due to either a crash failure or a network partition. 
If a newly-joined node discovers that it has restored quorum to the SL's consensus service,
then it knows that the SL is recovering from a crash failure.
Thus, its (perhaps newly-elected) monitor
sends \textit{announcements} to monitors in peer SLs
proposing that it now be considered alive.
If a quorum of monitors, using a BFT consensus algorithm, agree to this,
then the SL's status is updated by other monitors in the BG.
Since every RCanopus cycle requires one round of BFT consensus within each BG,
this is synchronized with BFT consensus on transaction order.
The new SL is given all the state that it missed by one of its peers, 
which is the set of its missed transactions along with their quorum certificates.

Similarly, if a BG wishes to rejoins the system after a system-wide network partition,
it must wait for the next periodic global consensus.
At this time,
newly-live BGs recover their state from peer BGs by obtaining missing transactions 
(with corresponding BG quorum certificates), 
that they can verify using 
global quorum certificates for missed epochs.

\section{Safety and liveness}
In this section, we prove that RCanopus is always safe and live when the situation
permits. Specifically, we prove the following theorem:
\\
\newtheorem{theorem}{Theorem}
\begin{theorem}[Safety and liveness]
RCanopus provides the following guarantees:
\begin{enumerate}
\item \textbf{Safety:} At the end of every consensus cycle, all live nodes 
agree on the same order of write transactions from all clients.
\item \textbf{Liveness:} In a system where 
\begin{itemize}
    \item up to $f$ crash failures can occur in each SL
    \item up to $f_i$ SL Byzantine failures can occur in $BG_i$
    \item up to $f_g$ BG leaders can have Byzantine failures 
\end{itemize}{}
if 
\begin{itemize}
    \item the number of nodes in each SL exceeds $2f$
    \item the number of SLs in $BG_i$ exceeds $3f_i$
    \item the number of BGs exceeds $3f_g$
    \item there is no system-wide network partition
\end{itemize}{}
then
every live node completes every consensus cycle, despite up to $f$ crash failures in every SL,
Byzantine failures in up to 
$f_i$ SLs in $BG_i$; up to $f_g$ Byzantine-failed BG leaders;
and up to $f_g$ BG crash failures.

If there is a partition, SLs/BGs in the super-majority partition, if such 
a super-majority partition exists, will be live and other SLs/BGs stall.
\end{enumerate}
\begin{proof} (sketch)
The proof is in two parts. 
First, in Section \ref{sec:cat}, we enumerate all possible faults in the system. 
Then, in Sections \ref{sec:mitnf}-\ref{sec:earlyexit}
we consider the impact of each fault on safety and liveness. We show that despite
faults, safety and liveness are lost only under the conditions enumerated in the statement of the theorem.
\end{proof}
\end{theorem}

\section{Fault categories}
\label{sec:cat}
This section enumerates the potential faults that may occur
in the RCanopus system and their potential impacts.
Over and above standard message failures, 
such as losses, duplication, and corruption,
faults arise from three causes: crashes, Byzantine failure, 
and network partition and they can affect either a node, an SL, a Byzantine group,
or the entire system\footnote{We ignore message failures based on our assumption of reliable
communication channels.}. 
Hence, we can enumerate all fault categories as follows:

\newcounter f
\addtocounter{f}{1}

\begin{table}[tbh]
\begin{center}
    \begin{tabular}{|l||l|l|l|}
    \hline
 \textbf{Entity} & \textbf{Crash possible?} &\textbf{Partition} & \textbf{Byzantine failure?}\\ 
  \ &  &\textbf{possible?} &  \\ \hline
    Node & Yes & N/A & Yes  \\ \hline
    SL & Yes&  Yes &  Yes\\ \hline
    $BG_i$ with $>3f_i SLs$& If $< f_i$ failures then no& Yes  & If $< f_i$ failures then no\\   &else yes & &else yes\\ \hline
     Global membership svc. & If $< f_g$ BG leader failures & Yes  & If $< f_g$ BG leader failures then no\\  with $>3f_g$ BG leaders &then no else yes & &else yes\\ \hline

    System & N/A & Yes  & N/A \\ \hline

    \end{tabular}
\end{center}
\caption {A broad categorization of all possible faults in the system. ``N/A" indicates the fault is either not possible or ignored.}
\end{table}

\textbf{Nodes} can fail by \textit{crashing}. If nodes fail during the first round of a cycle before
they share their state with at least one peer, their state will not be accessible
by other nodes and these transactions are lost to the system.
However, if they are able to communicate their transactions to at least one other node before failing,
these transactions may become available to the rest of the system. 
Node failures are masked by their SL peers, unless there are too many node
failures within an SL, 
in which case the SL itself fails.
Nodes can also launch several types of \textit{Byzantine} attacks.
The failure of a node that acts as a representative or a monitor has other consequences,
depending on the nature of the role, as discussed in more detail below. 

If too many nodes in an \textbf{SL} fail, then the SL loses quorum, and the entire
SL is stalled. This manifests itself as a SL \textit{crash} failure.
It is unlikely, but possible, that an SL \textit{partitions}. This may also result in
loss of quorum in the minority partition or both partitions. 
If so, the net result is either a set of node failures (similar to the node case above)
or an SL crash failure.
In case of partition, to prevent loss of safety due to a split brain, 
the minority partition must stall.
Moreover, to preserve liveness with safety, the majority partition
needs to achieve consensus on the set of failed/unreachable nodes.
Finally, SLs can manifest \textit{Byzantine} failures, and these are masked by their Byzantine group.

By definition, a \textbf{Byzantine group} $BG_i$with more than $3f_i+1$ SL
members is resilient to
Byzantine failures of up to $f_i$ SLs.
If there are more than $f_i$ failures,
to preserve safety, the entire BG should stall, causing a BG \textit{crash} failure.
As with an SL,
in case of \textit{partition}, to prevent loss of safety due to a split brain, 
the minority partition must stall.
Moreover, to preserve liveness with safety, the majority partition
needs to achieve consensus on the set of failed/unreachable SLs.
We use the same BFT consensus protocol both for consensus on transaction order
and on membership.

The \textbf{global membership service} is provided by the BG leaders. It also runs a
BFT consensus protocol, hence has the same failure conditions as individual BGs,
except that the number of failed BGs must be no more than $f_g$.

The entire \textbf{system} is susceptible to \textit{partition} failure.
These are due to crashes of multiple BGs or a network partition.
These failures need to be handled by the surviving BGs, if possible.
As with SLs and BGs,
in case of partition, to prevent loss of safety due to a split brain, 
the minority partition must stall.
Moreover, to preserve liveness with safety, the majority partition
needs to achieve consensus on the set of failed/unreachable BGs/SLs.

Finally, we need to deal with a special case, where a subtree of the system
subtended by the root completes a cycle, then fails, before communicating
with the rest of the system. This can cause a violation
of safety.

Given this discussion, we enumerate the potential set of faults in the system
in Table \ref{tab:list}.
In the next several sections, 
we discuss the impact of each class of failure on system safety and liveness.

\setcounter{f}{1}
\begin{table}[tbh]
\begin{center}
    \begin{tabular}{|l||l|l|}
    \hline
   \textbf{Failure class} & \textbf{Fault} & \textbf{Potential impact} \\
    \hline
    \textit{F\arabic{f}: Node crash\addtocounter{f}{1}} & Node failure in first round & Safety: inconsistent views on failed node's state\\ \hline
    \textit{F\arabic{f}: Node crash\addtocounter{f}{1}} & Emulator failure & Liveness: no response to representative \\ \hline
    \textit{F\arabic{f}: Node crash\addtocounter{f}{1}} & Representative failure & Liveness: SL missing state updates \\ \hline     
    \textit{F\arabic{f}: Node crash\addtocounter{f}{1}} & Monitor failure & SL may lose liveness \\ \hline
     \textit{F\arabic{f}: Node Byzantine\addtocounter{f}{1}} & Attack on transaction numbering &Safety: Transactions from this node \\&& likely to be first or last in global order\\ \hline
      \textit{F\arabic{f}: Node Byzantine\addtocounter{f}{1}} & Emulators non-responsive & Liveness: Same as SL crash failure\\ \hline   
                     \textit{F\arabic{f}: Node Byzantine\addtocounter{f}{1}} & Node ignores client &Safety: Client DoS \\ \hline
     \textit{F\arabic{f}: Node Byzantine\addtocounter{f}{1}} & Emulator lying & Safety: Any message from emulator may be a lie \\\hline

     \textit{F\arabic{f}: SL crash\addtocounter{f}{1}} & SL failure  & Safety: Unable to learn SL's state  \\&& Liveness: Peer SLs may stall \\ \hline

    \textit{F\arabic{f}: SL partition\addtocounter{f}{1}} & SL splits into partitions  & Safety: Possibility of split brain;\\&& Liveness: No SL partition may have quorum,\\&& leading to nodes stalling \\ \hline
   \textit{F\arabic{f}: SL Byzantine\addtocounter{f}{1}} & Messages from SL not trustworthy  & Safety: Cannot rely on any message\\&& from the failed SL\\ \hline
     \textit{F\arabic{f}: BG crash\addtocounter{f}{1}} & BG fails or is attacked  & Safety: Unable to learn BG's state\\&& Liveness: peer BGs may stall\\ \hline
          \textit{F\arabic{f}: BG partition\addtocounter{f}{1}} & SLs on different sides cannot & Safety: Inconsistent views of SL's state \\
     & communicate & \\ \hline
          \textit{F\arabic{f}: BG leader failure\addtocounter{f}{1}} & BG leader fails & Liveness: monitors and representatives stall\\ \hline
     \textit{F\arabic{f}: System partition\addtocounter{f}{1}} & BGs on opposite sides cannot & Safety: Inconsistent views of BG's state\\
     & communicate & \\ \hline  
     \textit{F\arabic{f}: Early exit\addtocounter{f}{1}} & A subtree of BGs  subtended by the & Loss of safety\\ 
     & root completes a cycle and fails&\\ \hline
    \end{tabular}
\end{center}
\caption {Potential faults.}
\label{tab:list}
\end{table}

\setcounter{f}{1}


\section{Impact of node failures on safety and liveness}
\label{sec:mitnf}
Note that, independent of its type,
one outcome of a node failure is to cause the eventual
removal of its corresponding ephemeral znode in its local ZooKeeper. 
The node's peers, who have access to this ZooKeeper service, can thus achieve consensus 
on its state.

\subsection{F\arabic{f}: Node failures during the first round}
\label{sec:first}
In the first round of each consensus cycle, 
nodes share their state (i.e., list of client transactions) with each other using atomic broadcast.
For safety, all nodes in the SL need to 
agree on the same set of transactions. 
By using an intra-SL consensus protocol, agreement on this set can be reached in an SL with more than 2$f$ nodes
despite up to $f$ failures. 

It is important to ensure that all nodes in the SL agree on the set of 
transactions \textit{before} these are shared with other SLs in the BG. 
Recall that transactions within an SL are shared only $b$ seconds after the start of the round,
and that the duration of a round is much larger than the
time taken to establish consensus within an SL (see Figure \ref{fig:pipeline}).
This procedure results in safety despite node failures.
Liveness of the SL is achieved as long as there are a
sufficient number of live
nodes within an SL to achieve quorum for the intra-SL consensus protocol. 

\subsection{Node failures during the second round} 

In the second round of each cycle,
monitors use a BFT consensus protocol
to achieve consensus as discussed in Section \ref{sec:byz}.
Such a protocol is tolerant of node failures, treating them similar to Byzantine failures.
Specifically, as long as there are enough live nodes and SLs to obtain a quorum
certificate, the BG maintains both safety and liveness
despite multiple node and SL faults\footnote{Of course, network partitions can cause
the BG to lose liveness. This is discussed later.}, else there is no liveness. 

\subsection{Node failures during subsequent rounds} 
We need to deal with the possibility of a node failure causing either an emulator or
a representative to fail. We deal with each case in turn. 

\addtocounter{f}{1}
\subsubsection{F\arabic{f}: Emulator crash failure}

A representative that relies
on an emulator that has crashed is able to make progress by attempting to obtain state from one of the crashes emulator's
peers, maintaining liveness. An SL only stalls if all emulators of a vnode 
fail, which corresponds to the failure of one or more BGs.
The impact of such a failure on liveness is discussed in Section \ref{sec:BGcrash}.

\addtocounter{f}{1}

\subsubsection{F\arabic{f}: Representative crash failure}

If one of the representatives in an SL fails, 
this failure is known the the SL's monitor, which asks another node in the SL
to become a representative, and takes over the responsibility 
of fetching remote state from the failed representative,
maintaining liveness.
Note that if the number of live nodes in the SL 
is $\le k$ the SL may have fewer than $k$ representatives. 
If all the representatives in an SL fail, this indicates that the SL
itself has failed, similar to an SL crash failure. The impact of this
failure on liveness is discussed in Section \ref{sec:slcrash}.

\addtocounter{f}{1}

\subsection{F\arabic{f}: Monitor failure}

If the SL's monitor fails, 
this is detected during the next intra-SL atomic broadcast by the other nodes in the SL.
On detecting this, 
a new leader is elected and asked to become the monitor. 
A newly-elected monitor also joins the appropriate BG. 



\addtocounter{f}{1}
\subsection{F\arabic{f}: Byzantine attack on transaction numbering}

With the Merkle-root based computation,
neither Byzantine nodes nor clients can game the system. Transaction ordering is completely dependent on other transactions in the current cycle. 
Furthermore, the information is hidden until the time of aggregation, 
which happens in parallel with batch sorting. Lastly, there is no overhead involved in this procedure, because the Merkle root is needed for the blockchain verification anyway.

\addtocounter{f}{1}
\subsection{F\arabic{f}: Byzantine attack: emulators non-responsive}
A Byzantine node serving as an emulator may not respond
to representatives. If this happens, then, from the representative's perspective,
the situation is identical to node failure, and the mitigation approach is also the same.


\addtocounter{f}{1}
\subsection{F\arabic{f}: Byzantine node ignores client}
\label{sec:dup}
A failed node can only ignore messages from its clients; it cannot
create fraudulent transactions on behalf of its clients because it does not have the 
client's private key.
To prevent transaction loss, recall that a timed-out client sends its transactions to $f_i$ additional
nodes in distinct SLs in the same BG.
   
\addtocounter{f}{1}
\subsection{F\arabic{f}: Byzantine emulator lies}
Due to the potential for Byzantine failures, any message coming from an emulator
is inherently unreliable.
To mitigate this problem,
every message from every emulator is accompanied by a quorum certificate
and is checked by the receiving representative.

\section{Impact of SL failures on safety and liveness}
\label{sec:mitslf}


\addtocounter{f}{1}
\subsection{F\arabic{f}: SL crash failures}
\label{sec:slcrash}

 It is possible that enough nodes  in the SL fail that there is 
  a loss of quorum in its  ZooKeeper service.
  In this case, the SL  will suffer from a crash failure,
  and, to preserve safety, all live nodes in the SL
should stall. This will appear to other SLs in the BG
as a network partition. We discuss how this is handled in Section \ref{sec:bgpart}, which deals
with BG partitions.

\addtocounter{f}{1}
\subsection{F\arabic{f}: SL partition}
\label{sec:slpartition}
An SL typically is comprised of servers on the same rack and connected by 
a single switch. Hence, it is very unlikely to partition.
Nevertheless, for the sake of completeness, we consider the impact of the partitioning of an SL
on system safety and liveness.
Two cases are possible: one of the SL partitions has a quorum of 
servers (i.e., it is the majority partition), or no SL partition has a quorum of
servers.

If there is no majority partition,
all nodes in the SL stall, and the situation is identical 
to the SL crashing (Section \ref{sec:slcrash}),
since stalled emulators will not respond to state requests when quorum is lost
(even with a null response).

If there is a majority partition, then, by definition,  
the majority partition has a quorum of  servers.
 These servers  establish consensus that all nodes in the minority partitions  are unreachable.
When a node is able to re-establish connection with the consensus service,
it should rejoin the SL using the join protocol in Section \ref{sec:join};
other nodes in the SL should not respond to excluded nodes.

\addtocounter{f}{1}
\subsection{F\arabic{f}: SL Byzantine failure}
\label{sec:byz}

It is possible for a Byzantine SL to attack the other SLs in its BG by:
\begin{enumerate}
\item giving inconsistent responses
to other SLs when asked for \textit{transactions} that it received from
its clients, which form its own state
\item lying about \textit{membership} in its own SL
\end{enumerate}

We consider each in turn.

\subsubsection{Byzantine attacks on transactions}

Recall that each client submits its transactions (eventually) to $f_i+1$ SLs.
Moreover, these transactions are signed with the client's  
private key and the set of transactions are ordered by
a hash on the transaction.
In the second round of each cycle,
these $f_i+1$ SLs use a BFT consensus algorithm to compute a
consistent order of client transactions (Section \ref{sec:bgs}).
Thus, even with $f_i$ failures,
a malicious SL cannot tamper with, reorder, or create fraudulent 
client transactions.
This implies that, at the end of  round 2 in each cycle,
it is possible to obtain a Byzantine fault-tolerant 
ordering of the write transactions.
along with the quorum certificate, mitigating Byzantine attacks on transactions.

In subsequent rounds, 
state queries received by emulators for their ancestor vnodes
are responded to with the state computed using the BFT algorithm,
along with a quorum certificate computed for each SL.
Thus, in subsequent rounds, the only possible attack by a malicious node would be 
to omit the transactions from one or more Byzantine groups. 
However, there can be at most $f_i$ malicious SLs in Byzantine group $i$.
Recall that representatives from other BGs (eventually) contact at least 
$f_i+1$ emulators (from $f_i+1$ different SLs) 
from each Byzantine group, so are guaranteed
at least one valid response. This prevents the attack. 

\subsubsection{Byzantine attacks on membership}
\label{sec:byzm}

Within a BG, the only Byzantine attack on membership possible
is for a malicious SL to lie about the nodes in its SL when participating in the BG
membership service (Section \ref{sec:bgs}).
It could pretend that it has some nodes that it doesn't actually have,
or not list all of its members.
However, there is no advantage to lying.
If the SL claims a node to be a member, but it is not one in reality,
then requests to this node will not result in responses.
On the other hand, if the SL does not reveal one of its nodes, then this node
will not be available to use as an emulator. In either case, the SL can, at worst,
deny certain requests, but cannot cause harm. 
%
%

\section{Impact of BG failures on safety and liveness}
\label{sec:mitbgf}
In this section, we discuss 
how to deal with more than $f_i$ SL crash failures in $BG_i$, 
stalling its BFT consensus protocol,
or when the BG partitions.

\addtocounter{f}{1}
\subsection{F\arabic{f}: BG crash failure}
\label{sec:BGcrash}
 It is possible that enough SLs  in the BG fail that there is 
  a loss of quorum in the BG  membership service.
  In this case, the BG  will fail.
  Live nodes in the BG detect  this as a loss of the BG's BFT consensus service.
In this case, to preserve safety, all live nodes
should stall. This will appear to other BGs in the system
as a network partition, and handled accordingly, as discussed in Section \ref{sec:sysf}.

\addtocounter{f}{1}
\subsection{F\arabic{f}: BG partition}
\label{sec:bgpart}

RCanopus tolerates a BG partition (or, equivalently, SL crash failure) by design.
Recall that the monitors of SLs in a BG need to achieve BFT consensus in each cycle. 
During a network partition event, 
the SL monitors in the super-majority partition (if one exists) will view
inaccessible SLs as having failed, and will create a quorum
certificate that only includes transactions from nodes in the SLs in the super-majority partition.
Monitors, and hence nodes, in the minority partition(s) stall, lacking the quorum to achieve 
BFT consensus.

When the partition heals, all nodes in the minority partition(s) will need 
to explicitly rejoin the system, and nodes in the super-majority partition should ignore messages
received from these nodes until they have explicitly rejoined the system.

Note that if the BG partitions such that all partitions have less than a super-majority
of SLs, then the BG loses liveness.

\addtocounter{f}{1}
\subsection{F\arabic{f}: BG leader failure}
\label{sec:bgpart}

BG leaders can fail by crashing or with a Byzantine fault. However, if fewer than $f_g$ leaders fail,
the use of BFT consensus amongst BG leaders assures safety at all times, and liveness when situations permit. 

\addtocounter{f}{1}
\section{F\arabic{f}: Impact of system partition on safety and liveness}
\label{sec:sysf}

We now discuss the case where a network partition event
separates some set of BGs from others.
Such a partition is nearly identical to the crash failure of one or more BGs
in that,
from the perspective of the other BGs,
some BGs are (perhaps temporarily) unreachable and therefore potentially dead.
What is important, however, is that nodes in BG that are actually alive,
but in a minority partition
must stall to avoid having a `split brain' situation, compromising safety.

When using the CM to deal with partition tolerance, 
we show in the Appendix, that using the CM leads to safety always, and liveness when possible.



The alternative is to use the global consensus service to deal with partitions.
Recall that BG leaders use a BFT consensus protocol to update their view of global
BG membership when monitors in a BG come to BFT consensus about the inaccessibility of
a remote BG.
This consensus is possible only in a super-majority partition,
thus stalled representatives and nodes can be unstalled only in such a partition.
Thus, all nodes in minority partitions will stall (guaranteeing safety) and nodes
in the super-majority partition (should one exist) will eventually 
achieve liveness.
Note that if the system partitions such that all partitions have less than a super-majority
of BGs, then the system loses liveness.

\addtocounter{f}{1}
\section{F\arabic{f}: Early exit}
\label{sec:earlyexit}

It is easiest to understand the problem of early exit
from a concrete example.
Consider the situation where the RCanopus system consists of four BGs subtended by the root.
Suppose that one of the BGs obtains 
all the information it needs from the other BGs, 
sends information back to its clients, then fails, but before
communicating any data to the other BGs.
In this case, the remaining BGs will exclude the failed BG's transactions,
but clients can receive inconsistent results, compromising safety.
This situation holds whenever a subtree of BGs subtended by the root all fail 
(or are partitioned) before communicating with the remainder. 

To prevent this, recall that nodes only commit
state computed in cycle $c$ at the end of their cycle $c+1$.
To see this this preserves safety,
note that for a BG $n$ to reach the end of its $c+1$ cycle,
it must receive input from every other live BG $m$ earlier in that cycle.
But this is only possible if $m$ was live at the end of the $c$th cycle.
Thus, if $n$ has reached the end of the $c+1$th cycle,
it (and all other live BGs who have reached the end of that cycle)
can be sure that $m$ was live at the end of the $c$th cycle,
which means that $m$'s contribution to global state in cycle $c$ can be safely committed.

Note that if $m$ did indeed fail then $n$ will stall before it reaches the end of the
$c+1$th cycle, triggering a BFT consensus on excluding $m$ in the $c+1$th cycle. 
In this case, all BGs need to achieve consensus on removing $m$'s state from the system
to achieve liveness with safety.
However, we expect this to happen rarely, so the recovery mechanisms can be somewhat complex.
In contrast, in the common case, consensus on safety is achieved with only a slight delay
and no loss of throughput. 
In short, a one-cycle-delay allows a node to establish consensus about safety without
having to consult an additional out-of-band consensus algorithm.

\section{Summary of mitigation mechanisms}
\label{sec:summary}
Table \ref{tab:list2} summarizes the mitigation mechanisms presented in this document.

\setcounter{f}{1}
\begin{table}[tbh]
\begin{center}
    \begin{tabular}{|l||l|l|}
    \hline
   \textbf{Failure class} & \textbf{Fault} & \textbf{Mitigation} \\
    \hline
    \textit{F\arabic{f}: Node crash\addtocounter{f}{1}} & Node failure in first round & Atomic broadcast, batch timers, one-cycle-delay\\ \hline
    \textit{F\arabic{f}: Node crash\addtocounter{f}{1}} & Emulator failure & global membership service updates emulator list; \\ && On timeout, retry after a $retry\_timeout$\\ \hline
    \textit{F\arabic{f}: Node crash\addtocounter{f}{1}} & Representative failure & Promote live node \\ \hline     
     \textit{F\arabic{f}: Node crash\addtocounter{f}{1}} & Monitor failure & Another node promoted to monitor status\\ \hline
          \textit{F\arabic{f}: Node Byzantine\addtocounter{f}{1}} & Attack on transaction numbering & Random nonce computed as hash on transaction\\&& BG orders on nonces\\ \hline
      \textit{F\arabic{f}: Node Byzantine\addtocounter{f}{1}} & Emulators non-responsive & Global membership service updates emulator list;  \\ && On timeout, retry after a $retry\_timeout$\\ \hline   
           \textit{F\arabic{f}: Node Byzantine\addtocounter{f}{1}} & Node launches DoS  attack& Standard DoS defenses\\ \hline
                     \textit{F\arabic{f}: Node Byzantine\addtocounter{f}{1}} & Node ignores client & Client sends $f+1$ transactions to distinct SLs \\ \hline
     \textit{F\arabic{f}: Node Byzantine\addtocounter{f}{1}} & Emulator lying & All emulator responses must be accompanied \\&& by quorum certificate \\ \hline
     \textit{F\arabic{f}: SL crash\addtocounter{f}{1}} & SL failure  & BGs tolerate up to $f$ SL \\&&  crashes due to the use of BFT consensus\\ \hline
    \textit{F\arabic{f}: SL partition\addtocounter{f}{1}} & SL splits into partitions  & Access to quorum gives \\ && majority liveness but minority stalls\\ \hline
   \textit{F\arabic{f}: SL Byzantine\addtocounter{f}{1}} & Messages from SL not trustworthy  &  All emulator responses must be accompanied\\&& by quorum certificate \\ \hline
     \textit{F\arabic{f}: BG crash\addtocounter{f}{1}} & BG fails  & 
Global consensus on the exclusion of the failed BG  \\ \hline
\textit{F\arabic{f}: BG partition\addtocounter{f}{1}} & SLs on different sides cannot & BGs tolerate partition due to\\&communicate & the use of BFT consensus within the BG \\
     &  & Minority partition stalls; super-majority is live\\ \hline
\textit{F\arabic{f}: BG leader failure\addtocounter{f}{1}} & BG leader crashed or Byzantine& Global BFT consensus \\ \hline
     \textit{F\arabic{f}: System partition\addtocounter{f}{1}} & BGs on opposite sides cannot &  
Consensus on the exclusion of BGs in minority \\&communicate &  partitions reached by BGs in supermajority\\ \hline
     \textit{F\arabic{f}: Early exit\addtocounter{f}{1}} & A subtree of BGs  subtended by the & One-cycle delay before commitment\\ 
     & root completes a cycle and fails&\\ \hline
    \end{tabular}
\end{center}
\caption {Mitigation techniques}
\label{tab:list2}
\end{table}

\section{Conclusion}
This document presents the design of the RCanopus system, which maintains several
essential aspects of the Canopus protocol, but adds several mechanisms to make it resilient
to a variety of faults. A detailed analysis of potential faults shows that RCanopus
can tolerate Byzantine failure, network partitioning, node crashes, and message loss.
We how how this can be achieved without sacrificing pipelining and the Canopus round-based
massively-parallel communication pattern.

In future work, we plan to implement and test RCanopus. We also plan to integrate into 
well-known permissioned blockchain systems such as HyperLedger Fabric and Parity's Substrate. 

\bibliography{rf}
\bibliographystyle{acm}

\appendix

\section{The Convergence Module (CM)}\label{app:converg}
\label{sec:cmapp}
The Convergence Module (CM) is a mechanism for ensuring convergence in all-to-all communication despite network partitions and failures of components.
It offers the following advantages:
\begin{enumerate}
\item It maintains safety in an asynchronous environment.
\item An administrator-defined policy can be defined to determine how to deal with transactions proposed by unavailable BGs.
\end{enumerate}
The high-level approach is applicable to all three layers of RCanopus (intra-SL, inter-SL/intra-BG, and inter-BG) but for concreteness this section describes specifically the variation applicable to the inter-BG layer, where it is implemented using a Byzantine fault-tolerant replicated state machine (RSM) 
deployed on the same physical infrastructure as the BGs.
For example, one replica of the CM can be hosted in each BG as long as the total 
number of BGs is at least $3f + 1$ where $f$ is an upper bound 
on the number of simultaneous replica failures.

More concretely, the CM service is a collection of \emph{CM nodes}, each comprising an RSM replica for fault-tolerant distributed coordination, and a remote procedure call (RPC) server for interaction with BGs.
The RPC server is able to read the state of the RSM replica, but cannot write it directly.
State changes are instead accomplished by submitting commands from the service handler of the RPC server to the RSM by way of the co-located RSM replica.

One of the difficulties with using replicated state machines and consensus in a Byzantine environment is that Byzantine processes can in some cases propose invalid commands or inputs.
That is, even though agreement is guaranteed on some decision or sequence of decisions, these decisions themselves may be invalid because they are based on erroneous or malicious inputs.
We deal with this issue using a combination of techniques that ensure the following guarantees with respect to decisions made using the intra-BG BFT consensus and the CM replicated state machine:
\begin{enumerate}
\item Integrity: each decision was agreed to by a sufficiently large quorum of processes (i.e., a supermajority).
\item Validity: each decision is consistent with the protocol, regardless of whether it was proposed by an honest node or a Byzantine node.
\end{enumerate}
The correctness properties are ensured using various forms of certificates, which typically contain a collection of signatures from quorum of $f+1$ nodes.
Details are provided in Section~\ref{app:cert}.

\newcommand{\BGR}{\texttt{BG\_REPORT}}
We begin by describing a simplified version of the protocol in which the BGs communicate with the CM (i.e., with the local CM node) when they approach the end of round 3 of every consensus cycle.
Specifically, this occurs when the BG has received inputs from $2f+1$ BGs including itself, and has either received inputs from or timed out on the remaining $f$ BGs.
Each BG sends a \BGR\ message to all the CM replicas indicating the set of other BGs from which transactions were successfully received up to this point in the current cycle, as well as a hash of the transactions for each peer BG.

For each cycle, a CM node waits until it has received \BGR\ messages from $2f+1$ of the BGs, and then waits further until each remaining BG either sends its message, or is suspected to have failed due to a timeout.\footnote{In this case ``failed'' means unreachable rather than crashed.}
Suspicions of failure are corroborated by other CM nodes as follows to prevent false alarms by Byzantine nodes:
failure is declared when a quorum of $f+1$ CM nodes all suspect (i.e., have timed out on) a particular BG.
Messages relaying failure suspicion are signed by CM nodes and contain both the BG's ID and the cycle number to prevent replay attacks.
A collection of $f+1$ such signatures from distinct CM nodes for the same BG and cycle number comprise a \emph{failure certificate}.
If a failure certificate cannot be computed for some BG, then the CM node eventually receives a \BGR\ for that BG indirectly from some other CM node.

As a running example, consider the case of three BGs: BG1, BG2, and BG3.  
Suppose that BG1 in cycle C1 receives inputs from BG2 and BG3.  If BG2 becomes unreachable in cycle C1 with respect to BG3 and the CM, then the CM node might receive the following two \BGR\ messages from BG1 and BG3, respectively:
\begin{quote}
	[C1, BG1:hash1, \{BG2:hash2, BG3:hash3\}]
    \end{quote}
    \begin{quote}
	[C1, BG3:hash3, \{BG1:hash1\}]
\end{quote}
\noindent The \BGR\ is called \emph{complete} in the first case, indicating that the BG has received transaction inputs from all BGs, and \emph{incomplete} otherwise.

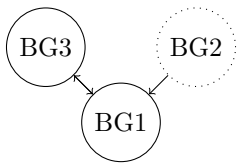
\begin{figure}[h]
\centering
\begin{tikzpicture}
    \node[shape=circle,draw=black] (BG1) at (1,0) {BG1};
	\node[shape=circle,draw=black,dotted] (BG2) at (2,1) {BG2};
	\node[shape=circle,draw=black] (BG3) at (0,1) {BG3};
    
    \path [->](BG2) edge node[left] {} (BG1);
    \path [->](BG3) edge node[left] {} (BG1);
    \path [->](BG1) edge node[left] {} (BG3);
\end{tikzpicture}
\caption{Graph representation of CM's inputs for cycle C1.}\label{fig:graphc1}
\end{figure}

Suppose that a CM node eventually computes a failure certificate for BG2, and creates a graph representation of cycle C1 as shown in Figure~\ref{fig:graphc1}.
The vertices of the graph represent BGs and the edges indicate that one BG has a copy of another BG's transactions.
The direction of the edge is from the BG that issued the transactions to the BG that received a copy of the transactions.

\newcommand{\CMR}{\texttt{CM\_REPLY}}
Next, the CM node computes the outcome of C1 using a configurable policy.
In one variation, it computes a maximum subset of vertices with out-degree greater than or equal to $N-1$, where $N$ is the total number of non-failed vertices (i.e., vertices representing live BGs).
Such vertices represent proposals that are fully replicated across all non-failed BGs.
For C1, $N = 2$ because BG2 is unavailable, which is indicated in Figure~\ref{fig:graphc1} using a dotted circle around BG2.
Any such vertex is implicitly excluded from the computation, along with any edges incident on it.
The outcome of the CM's graph analysis in C1 is therefore the maximum subset of non-failed vertices with in-degree one: \{BG1, BG3\}.
The remaining BG, namely BG2, is then added to the set of faulty nodes (FN) for cycle C1.
This means that BG2's inputs are excluded from C1, and does not imply BG2's deletion from the group membership of the system.
Next, the CM node proposes a command to the RSM that associates two records with cycle C1: the mapping \{BG1:hash1, BG3:hash3\} representing the collection of transactions committed in cycle C1, and the mapping FN = \{BG2:failcert2\} representing the BGs deemed to have failed and their failure certificates.
Each CM node eventually receives the command, and is able to report to its own BG the outcome of cycle C1.
The CM node finally delivers to the BG a \CMR\ message that contains the command and corresponding certificate (see Section~\ref{app:cert}).
A BG that receives such a message from the local CM node will stop waiting for any BG in FN, and continues to contact other BGs to retrieve any missing transaction inputs.

The graph analysis is performed in parallel by different CM nodes for each cycle, and this may lead to differing views on the liveness of a particular BG, hence to different outputs.
The certification scheme described in Section~\ref{app:cert} ensures that the decision of the CM is well-defined for each cycle despite this.
Specifically, in the event that multiple decisions are committed in the CM RSM, the first state transition is treated as authoritative and the others are ignored.
Duplicate decisions should nevertheless be avoided in the interest of performance, and several optimizations can be used for this purpose:
\begin{itemize}
\item The CM nodes submit a command to the RSM for cycle $C$ only if they have not yet executed a state transition command for $C$.
\item Randomized timeouts can be used prior to submitting an RSM command.
\item A distinguished leader CM node can be responsible for submitting the RSM command, in which case duplicate state transitions for the same cycle would occur only if the leader suffers a Byzantine failure or if two leaders exist temporarily because of inaccurate failure detection.
A RAFT-style term-based leader election algorithm can be used in this context with extensions for Byzantine fault tolerance (e.g., a node cannot start leader election until is has computed a failure certificate for the current leader).
\end{itemize}

\remove{
Another technicality arises if a BG suffers more than its tolerated number of Byzantine failures, in which case different peer BGs can receive different versions of the failed BG's transaction inputs.
Such a failure is detected by the presence of conflicting hashes received for the same source BG by the CM node from different destination BGs.
In this case the Byzantine BG is added to FN, and the conflicting hashes signed by two or more peer BGs comprise a failure certificate.
A compromised BG can also lead to a Byzantine failure of a CM node, which can submit bogus commands to the RSM.
Commands that are malformed or contain invalid signatures are ignored even if they are committed by the RSM.
The commands can be signed by the CM node that issues them (using a BG-specific key), in which case a Byzantine CM node can be identified easily.
}

Now suppose that BG2 was merely slow and not faulty, and continues to participate in the next cycle C2.  
Since BG2 is added to FN specifically for cycle C1, BG1 and BG3 continue to attempt communication with BG2 in cycle C2.\footnote{An optimization that avoids this is described later on in Section~\ref{sec:cmnetpart}.}
The CM node may therefore receive the following messages in C2:
\begin{quote}
	[C2, BG1:hash1, \{BG2:hash2, BG3:hash3\}]
    \end{quote}
    \begin{quote}
	[C2, BG2:hash2, \{BG1:hash1, BG3:hash3\}]
    \end{quote}
    \begin{quote}
	[C2, BG3:hash3, \{BG1:hash1, BG2:hash2\}]
\end{quote}
    
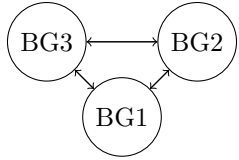
\begin{figure}[h]
\centering
\begin{tikzpicture}
    \node[shape=circle,draw=black] (BG1) at (1,0) {BG1};
    \node[shape=circle,draw=black] (BG2) at (2,1) {BG2};  
    \node[shape=circle,draw=black] (BG3) at (0,1) {BG3};
    
    \path [->](BG2) edge node[left] {} (BG1);
    \path [->](BG3) edge node[left] {} (BG1);
    \path [->](BG1) edge node[left] {} (BG2);
    \path [->](BG3) edge node[left] {} (BG2);
    \path [->](BG1) edge node[left] {} (BG3);
    \path [->](BG2) edge node[left] {} (BG3);
\end{tikzpicture}
\caption{Graph representation of CM's inputs for cycle C2.}\label{fig:graphc2}
\end{figure}

\noindent Figure~\ref{fig:graphc2} shows the corresponding graph computed by the CM.
The outcome of the graph computation for C2 is different from C1 because BG2 participates fully.
The CM node then associates \{BG1:hash1, BG2:hash2, BG3:hash3\} and FN = \{\} with cycle C2 by issuing a command
to the RSM.

Next, the CM may receive the following messages in cycle C3 if BG2 is removed from the BG by the administrator in cycle C2:
\begin{quote}
	[C3, BG1:hash1, \{BG3:hash3\}]
    \end{quote}
    \begin{quote}
	[C3, BG3:hash3, \{BG1:hash1\}]
\end{quote}

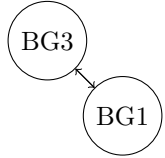
\begin{figure}[h]
\centering
\begin{tikzpicture}
    \node[shape=circle,draw=black] (BG1) at (1,0) {BG1};
    \node[shape=circle,draw=black] (BG3) at (0,1) {BG3};
    
    \path [->](BG3) edge node[left] {} (BG1);
    \path [->](BG1) edge node[left] {} (BG3);
\end{tikzpicture}
\caption{Graph representations of CM's inputs for cycle C3.}\label{fig:graphc3}
\end{figure}

\noindent Figure~\ref{fig:graphc3} shows the corresponding graph computed by the CM.
The CM then associates \{BG1:hash1, BG3:hash3\} and FN = \{\} with cycle C3 by issuing a command to the RSM, similarly to cycles C1 and C2.

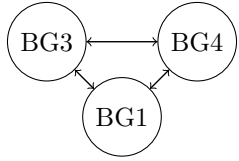
\begin{figure}[h]
\centering
\begin{tikzpicture}
    \node[shape=circle,draw=black] (BG1) at (1,0) {BG1};
    \node[shape=circle,draw=black] (BG4) at (2,1) {BG4};
    \node[shape=circle,draw=black] (BG3) at (0,1) {BG3};
    
    \path [->](BG3) edge node[left] {} (BG1);
    \path [->](BG4) edge node[left] {} (BG1);
    \path [->](BG1) edge node[left] {} (BG3);
    \path [->](BG4) edge node[left] {} (BG3);
    \path [->](BG1) edge node[left] {} (BG4);
    \path [->](BG3) edge node[left] {} (BG4);
\end{tikzpicture}
\caption{Graph representations of CM's inputs for cycle C4.}\label{fig:graphc4}
\end{figure}

Next, suppose that BG2 comes back as BG4 in cycle C4.  Then in cycle C4 the CM may receive the following messages:
\begin{quote}
	[C4, BG1:hash1, \{BG3:hash3, BG4:hash3\}]
    \end{quote}
    \begin{quote}
	[C4, BG3:hash3, \{BG1:hash1, BG2:hash2\}]
    \end{quote}
    \begin{quote}
	[C4, BG4:hash4, \{BG1:hash1, BG3:hash3\}]
\end{quote}
This scenario indicates a return to steady state operation in which each BG is able to communicate with every other BG.
Then the CM computes the graph shown in Figure~\ref{fig:graphc4} and associates \{BG1:hash1, BG3:hash3, BG4:hash4\} and FN = \{\} with cycle C4 using the RSM since all vertices have out-degree two.

In the running example above, the output of the RCanopus protocol for a given cycle is determined by the set of nodes recorded in the CM RSM.  For cycles C2 and C4, the output is the union of the inputs of all the BGs in the system.  For C1 and C3, the output is the union of the inputs of BG1 and BG3 only.

In a variation of the protocol, the graph analysis selects vertices whose out-degree is sufficient to achieve an administrator-defined replication factor $R$.
For example, to ensure a replication factor of at least $R$ under the assumption that a BG never crashes permanently, the CM would select a maximal subset of vertices with out-degree at least $R-1$.  (We assume that a BG stores its own inputs, hence only $R-1$ \emph{additional} copies are required at other BGs.)
In this optimized version of the protocol, the CM may associate additional information with each consensus cycle to identify the BGs that hold copies on the transaction inputs that constitute the output.



To summarize, the CM uses communication successes in a cycle, along with a graph algorithm, to determine the set of BGs that can be considered live for the cycle.
This information is broadcast to all the BGs, who then exclude transactions from any BGs that were identified by the CM as failed (meaning unavailable).
The cycle completes only at this point.
Section~\ref{app:cmbypass} discusses how the CM can be removed from the critical path in failure-free operation.

\subsection{Bypassing the CM in the absence of failures}\label{app:cmbypass}
In the interest of performance, the BGs must avoid coordinating with the CM in every cycle in the common case of failure-free operation.
The protocol can be modified to remove the CM from the critical path in this case, and invoked only when needed to settle the outcome of a consensus cycle that is affected by a failure.
Letting $k$ denote the depth of the processing pipeline, this is accomplished by having each BG broadcast meta-data related to the outcome of cycle $C$ to the other BGs as part of its input in cycle $C+k$, which begins only after cycle $C$ completes.\footnote{This optimization requires that a BG participate for one additional cycle after it is deleted from the system's group membership. In particular, a BG that is deleted in cycle $C$ must still offer its meta-data (but no transactions) in cycle $C+k$ before disengaging entirely from the protocol.}
At the completion of cycle $C+k$, each BG can retroactively analyze the meta-data received for cycle $C$ to classify this cycle as belonging to one of two disjoint categories:
\begin{enumerate}
\item A \emph{CM-assisted} cycle is one where at least one BG received assistance from the CM, in which case the outcome of the cycle is decided explicitly by the CM.\footnote{``Received assistance'' means that a BG reported to the CM that it is stalled on another peer BG, and was told the outcome of the cycle by the CM.}
\item A cycle is \emph{unassisted} otherwise, in which case its outcome is the union of the inputs of all the BGs in the system.
\end{enumerate}
We say that the commitment of cycle $C$ is \emph{delayed by $k$ cycles} because the outcome of cycle $C$ is not known until the end of cycle $C+k$.

\newcommand{\CMD}{\texttt{CM\_DENY}}
\newcommand{\BGM}{\texttt{BG\_META}}
\newcommand{\BGMNA}{\BGM\texttt{-NO\_ASST}}
\newcommand{\BGMA}{\BGM\texttt{-ASSISTED}}
\newcommand{\BGMD}{\BGM\texttt{-DENIED}}
As hinted earlier, the protocol is modified by including meta-data regarding cycle $C$ in the input for cycle $C+k$.
The meta-data is packaged as a \BGM\ message indicating whether or not this BG received assistance from the CM in cycle $C$.
The \BGM\ message comes in three flavors:
\begin{enumerate}
\item A \BGMNA\ message indicates that the BG decided not to seek assistance from the CM, and hence did not receive assistance.
\item A \BGMA\ message indicates that the BG sought and received assistance from the CM.
\item A \BGMD\ message indicates that the BG sought assistance from the CM and was denied because its request came too late (i.e., at a time when the CM had already begun assisting some BG with a later cycle).  This case is discussed in more detail shortly.
\end{enumerate}
To ensure a correct categorization of cycles, the meta-data reported by BGs must be in agreement with the state of the CM meaning that the cycle is CM-assisted if and only if at least one BG generates a \BGMA.  To that end, BGs must follow certain rules:
\begin{enumerate}
\item The decision of a BG to seek (or not seek) assistance from the CM is committed in a Byzantine fault tolerant manner and certified using the techniques described in Section~\ref{app:cert}.
For performance, this decision can be combined with consensus on the BG's input for some future cycle (e.g., cycle $C+k$).
The certificate corresponding to this decision is attached to the \BGM\ message.
\item Once a BG records a decision to seek assistance, it cannot complete the cycle without submitting a request to the CM and receiving a response.
The request message is an incomplete \BGR.
The response message is either a \CMR, which was discussed earlier, or a \CMD, which is explained shortly.
\item A \CMR\ message is attached to a \BGMA\ meta-data message to prove that a BG received assistance.
Similarly, a \CMD\ message is attached to a \BGMD\ meta-data message to prove that a BG was denied.
The certification scheme discussed in Section~\ref{app:cert} ensures that a BG can only obtain either a \CMR\ or \CMD\ for a given cycle, and not both.
\end{enumerate}


As an example, suppose that there are three BGs -- BG1, BG2, BG3 -- in a system with a pipeline of depth $k=1$.
Suppose further that the proposals of these three BGs are exchanged successfully in cycle C1 without the CM's assistance.
Then each BG adds a \BGMNA\ message to its input for cycle C2.
At the end of cycle C2, presuming no failures, each BG computes the following mapping from BGs to their meta-data:

	C1: \{BG1 $\rightarrow$ \BGMNA, BG2 $\rightarrow$ \BGMNA, BG3 $\rightarrow$ \BGMNA\}

\noindent The number of keys in this mapping (three), and the fact that all values are \BGMNA, indicate collectively that cycle C1 was unassisted.  All BGs know this once C2 is complete because in this case all BGs receive meta-data from all other BGs.

Next, consider a failure scenario in which BG1 is unable to receive data from BG2 in cycle C3, and requests (as well as receives) assistance from the CM in cycle C3.
Then BG1 includes a \BGMA\ message in its input to cycle C4.
The outcome of cycle C4 depends on the meta-data received by a given BG at the end of cycle C4.
If all the meta-data messages are \BGMNA\ or \BGMD, then C3 is unassisted (not the case in this example).
If at least one meta-data message is a \BGMA\ then C3 is CM-assisted.
For example, this occurs in the current example if a BG computes the following mapping for C3:

	C3: \{BG1 $\rightarrow$ \BGMA, BG2 $\rightarrow$ \BGMNA, BG3 $\rightarrow$ \BGMNA\}
    
\noindent The same case applies if a BG computes a subset of this mapping as follows:

	C3: \{BG1 $\rightarrow$ \BGMA, BG2 $\rightarrow$ \BGMNA\}

\noindent Finally, if there is one or more missing meta-data message, and no \BGMA\ is received, then further investigation is required to resolve the outcome of the cycle.
For example, this occurs if a BG computes the following mapping for C3 in C4:

	C3: \{BG2 $\rightarrow$ \BGMNA, BG3 $\rightarrow$ \BGMNA\}

\noindent In this case cycle C4 itself is CM-assisted because the input of BG1 for C4 is not received by the BG that computes the above mapping, and so the status of C3 (i.e., assisted vs.\ unassisted) is known already to the CM provided that it maintains the following invariant: the CM cannot assist with cycle $C$ if it has already begun assisting with cycle $C+k$ where $k$ is the depth of the pipeline.
Maintaining this invariant ensures that the CM can be queried directly to determine the outcome of C3 once C4 is known to be CM-assisted.
Enforcing the above invariant is difficult as it leads to a race condition in which a slow BG requests assistance for an earlier cycle $C$ once the CM has already started assisting a later cycle $C'$.
A CM node deals with this race condition as follows:
before replying to any request for assistance with $C'$, it ensures that a supermajority of BGs have progressed past cycle $C$ after either receiving assistance from the CM or receiving transaction inputs from all other BGs in cycle $C$.
More concretely, the CM waits until it either has a complete \BGR\ for cycle $C$ from a supermajority of BGs, or else until some BG issues an incomplete \BGR\ for cycle $C$, indicating a request for assistance.
In the latter case the CM offers assistance by returning a \CMR\ message. 
Any request for assistance (i.e., incomplete \BGR) with cycle $C$ received after this point is replied to with a precomputed \CMR\ if $C$ was CM-assisted, and with a \CMD\ message otherwise.
The denial message indicates to the requesting BG that $C$ is unassisted and any missing inputs can be obtained by querying any supermajority quorum of BGs.
\begin{theorem}[safety]
If two BGs determine the outcome of a cycle $C$ then either both decide that $C$ is CM-assisted or both decide that $C$ is unassisted.
\begin{proof}
Suppose for contradiction that two BGs, say BG1 and BG2, reach conflicting decisions regarding cycle $C$.
Suppose that BG1 decides that $C$ is CM-assisted, and BG2 decides that $C$ is unassisted.
Since BG1 decides that $C$ is CM-assisted, one of three events occurred:
\begin{enumerate}
\item[(i)] BG1 requested and received assistance from the CM in cycle $C$, and produced a \BGMA; or
\item[(ii)] BG1 received a \BGMA\ from another BG in a later cycle; or
\item[(iii)] BG1 received incomplete meta-data for cycle $C$ and determined that $C$ was CM-assisted by querying the CM directly (i.e., received a \CMR\ and not a \CMD).
\end{enumerate}
In cases (i) and (iii), it follows that BG1 reaches the correct conclusion since it communicates directly with the CM, and since the certification scheme described in Section~\ref{app:cert} ensure that the CM nodes produce a uniform decision for a given cycle.
In case (ii), the \BGMA\ includes a corresponding \CMR, which is certified and proves that the cycle was CM-assisted.

Next, consider how BG2 reached the incorrect conclusion that $C$ is unassisted.
According to the protocol, BG2 must have received only \BGMNA\ and \BGMD\ meta-data messages from all BGs in cycle $C+k$, and yet one of these BGs did in fact receive assistance in cycle $C$ (i.e., was offered a \CMR\ for cycle $C$).
A \BGMNA\ includes a certificate that proves the BG committed not to seek assistance from the CM in cycle $C$.
On the other hand, a \BGMD\ includes a \CMD\ message, which proves that the cycle was unassisted.
Both cases contradict the observation that the BG under consideration received assistance from the CM in cycle $C$.
\end{proof}
\end{theorem}

\begin{theorem}[liveness]
Suppose that the protocol begins to process cycle $C$ and that the pipeline depth is $k$.
Suppose further that each BG individually maintains safety and liveness.
If there exists a quorum $Q$ of at least $2f+1$ BGs that are able to exchange and process messages
sufficiently quickly to avoid triggering timeouts in the protocol, then
eventually every BG in $q$ progresses to a cycle $C \geq C + k$.
\begin{proof}
The protocol begins with each BG collecting inputs from peer BGs.
Every BG in $Q$ is able to receive such inputs from all other BGs in $Q$, and then either
receives inputs from or times out on every BG outside of $Q$.
If the BG receives inputs from all BGs then it proceeds directly to cycle $C + k$ with a \BGMNA\ 
meta-data message for cycle $C$, and the theorem holds.
On the other hand, if any of the inputs is missing then the BG requests assistance from the CM by sending a \BGR\ message to its local CM node.
(This case can be avoided if an unavailable BG is excluded by the CM for cycle $C$ while assisting with some earlier cycle, or temporarily removed from the group by the administrator.)
At this point the CM node follows one of several execution paths.

\emph{Case~A:} the CM node has already computed a \CMR\ or \CMD\ for cycle $C$.
Then the CM node returns the same response to the BG under consideration.

\emph{Case~B:} the CM node has not yet computed a \CMR\ or \CMD\ for cycle $C$.
In this case the CM node first ensures that the outcome of cycle $C-k$ has been determined,
which involves contacting a quorum of $2f+1$ BGs, analyzing their \BGR\ messages, and in some cases
computing one or more failure certificates.
Timely communication with BGS and corresponding CM nodes in $Q$ ensures that this part of the protocol
completes eventually.
Next, the CM node either attempts to assist the BG, or replies with a denial if it has already computed a certified \CMR\ (i.e., provided assistance) for cycle $C + k$ or higher.

\emph{Subcase~B1:} the CM node has already computed a certified \CMR\ for cycle $C + k$ or higher.
Then the CM node replies to the BG with a \CMR\ message if such a message has been processed by the RSM and certified for cycle $C$, or else it replies with a \CMD\ message, which is also be certified.
In the former case, the reply to the BG is pre-computed.
In the latter case, the \CMD\ command can be certified using the CM nodes corresponding to $Q$.

\emph{Subcase~B2:} the CM node has not yet computed a \CMR\ for cycle $C + k$ or higher.
Then the CM node proceeds with graph analysis based on the inputs of the BGs in $Q$ and possibly other BGs outside of $Q$.
Every BG in $Q$ reports its \BGR\ to the CM node, and for every BG outside of $Q$, either a failure certificate is computed by $2f+1$ CM nodes, or a \BGR\ message is retrieved for cycle $C$.
Next, the CM node proposes a command to its RSM based on the computed \CMR.
This command is accepted by the RSM, particularly the CM nodes corresponding to the BGs in $Q$,
but there is no guarantee that it will be certified because a conflicting command for cycle $C$
may have a already been proposed.
If the command is certified successfully, then the CM node returns a certified \CMR\ response to the BG.
Otherwise the CM node discovers that either a \CMR\ message was already processed for cycle $C$ or for cycle $C+k$ or higher in an earlier RSM command, and was successfully certified.
In the former case, the CM node replies with the earlier \CMR\ command after ensuring that the command is certified.
In the latter case, the CM replies with a \CMD\ message to the BG.
\end{proof}
\end{theorem}

\remove{
\paragraph{Case 1: An BG receives all three meta-data bits.}
Then this BG computes the following mapping

	C3: \{BG1 $\rightarrow$ 1, BG2 $\rightarrow$ 0, BG3 $\rightarrow$ 0\}

\noindent The presence of a 1 bit in the value set identifies C3 correctly as a CM-assisted cycle.

\paragraph{Case 2: An BG receives fewer than three meta-data bits.}
In this case C4 itself is CM-assisted, as otherwise it could not finish without the inputs of all BGs.
The outcome of C3 is determined by a closer inspection of the meta-data.

\paragraph{Case 2a: Some meta-data bit is 1.}
For example, the BG may compute the following mapping if BG3 failed in cycle C4:

	C3: \{BG1 $\rightarrow$ 1, BG2 $\rightarrow$ 0\}

\noindent The presence of a 1 bit in the value set identifies C3 correctly as a CM-assisted cycle, as in Case~1.

\paragraph{Case 2b: all meta-data bits are 0.}
For example, the BG may compute the following mapping if BG1 failed in cycle C4:

	C3: \{BG2 $\rightarrow$ 0, BG3 $\rightarrow$ 0\}

\noindent This is the most challenging case since the incomplete mapping makes it impossible to determine directly whether or not BG1 received assistance from the CM in cycle C3, and BG1 may not be available to report this information since it was suspected of failure.
}

\subsection{Optimization for long network partitions}\label{sec:cmnetpart}
A network partition may cause some BGs to lose contact with other BGs in the system for an extended period of time.
The CM bypass mechanism, as described earlier, is ineffective during such a partition because every cycle must be CM-assisted, which introduces a substantial performance overhead.
The problem can be remedied by excluding the inputs of BGs selected in the graph analysis for multiple consensus cycles.
That is, in a CM-assisted cycle $C$, the decision to exclude the inputs of some BG persists beyond cycle $C$ for one or more additional cycles starting at $C+k$, where $k$ is the depth of the pipeline.
The additional information regarding excluded BGs can be embedded in the \CMR\ message, which is attached to the \BGMA\ messages in the CM bypass mechanism.
An excluded BG resumes computation later on without having to change its ID and execute a join protocol.
Peer BGs continue to send messages to the excluded BG until then, but do not expect replies.
Alternatively, a temporary group membership change can be made using the same mechanism as for ordinary group membership, which remains available from inside a sufficiently large partition (i.e., as long as a supermajority of BGs can be reached).


\subsection{Certification of consensus decisions}\label{app:cert}
This section pertains to two types of decisions:
\begin{enumerate}
\item The BG's choice of transaction inputs for a given consensus cycle, which may include some meta-data regarding a past cycle (see Section~\ref{app:cmbypass}).
\item The CM's decision on the outcome of a cycle, which is based on graph analysis.
\end{enumerate}
Since both decisions are reached using a black box BFT consensus, it is possible to ensure integrity by computing a quorum certificate for a decision -- a collection of signatures from a supermajority ($2f+1$) quorum of servers.
However, the quorum certificate created in an application-agnostic manner inside the BFT black box does not by itself ensure validity in a Byzantine environment.
This is because an invalid decision may be committed based on the input of a Byzantine node.
Worse yet, multiple conflicting decisions may be committed.

One solution to the problem is to modify the implementation of the BFT consensus and filter out both invalid and duplicate decisions before they are committed.
In this case a conventional quorum certificate is sufficient to ensure both integrity and validity.
If the BFT implementation cannot be modified, for example because it is closed source or managed by a third party provider, then We solve the problem by applying additional signatures on top of the quorum certificate that attest to the fact that the decision is singular and intended by the protocol.
The number of additional signatures must be sufficient to ensure that at least one signature comes from an honest node (i.e., at least $f+1$).
This node must be aware of all prior decisions reached by the BFT consensus so that it can detect invalid or duplicate decisions.
An invalid decision must not be signed, and only the first decision in a sequence of duplicate decisions may be signed.

Example: A representative from a BG proposes that the BG's transaction input in the inter-BG layer for some cycle $C$ is a set $S$.
A second representative proposes a different set $S'$ (e.g., same transactions but different meta-data for an earlier cycle in the CM bypass optimization).
An honest server proposes $S$ to the intra-BG BFT consensus, and a Byzantine node subsequently proposes $S'$.
The BFT records both decisions -- $S$ followed by $S'$.
The additional signatures will ensure that $S$ is identified correctly as the BG's input for cycle $C$.
If the quorum certificates produced for $S$ and $S'$ include the index number of the corresponding decision, then any honest node identifies $S$ as the first valid decision computed by the BFT consensus for cycle $C$, signs $S$, and refuses to sign $S'$.
\end{document}